\documentclass[reprint,aps,prb,amssymb, amsmath,superscriptaddress]{revtex4-2}

\usepackage{graphicx}
\usepackage{pifont}
\usepackage{hyperref}
\hypersetup{
    colorlinks=true, 
    linktoc=all,     
    linkcolor=blue,  
    citecolor =blue
}
\usepackage[T1]{fontenc}
\newcommand{\bk}{{\bf k}}

\newcommand{\nn}{\nonumber}

\newcommand{\bq}{{\mathbf q}}

\begin{document}
\title{Magnetic hard-direction ordering in anisotropic Kondo systems}

\author{M. P. Kwasigroch}
\affiliation{Department of Mathematics, University College London, Gordon St., London, WC1H 0AY,  United Kingdom}
\affiliation{Trinity College, Trinity Street, Cambridge, CB2 1TQ, United Kingdom}
\author{Huanzhi Hu}
\affiliation{London Centre for Nanotechnology, University College London, Gordon St., London, WC1H 0AH, United Kingdom}
\author{F. Kr\"uger}
\affiliation{London Centre for Nanotechnology, University College London, Gordon St., London, WC1H 0AH, United Kingdom}
\affiliation{ISIS Facility, Rutherford Appleton Laboratory, Chilton, Didcot, Oxfordshire OX11 0QX, United Kingdom}
\author{A.~G. Green}
\affiliation{London Centre for Nanotechnology, University College London, Gordon St., London, WC1H 0AH, United Kingdom}

\date{\today}

\begin{abstract}
We present a generic mechanism that explains why many Kondo materials show magnetic ordering along directions
that are not favoured by the crystal-field anisotropy.  Using a renormalization-group (RG) analysis of single impurity Kondo models 
with single-ion anisotropy, we demonstrate that strong fluctuations above the Kondo temperature drive a moment re-orientation over 
a wide range of parameters, e.g. for different spin values $S$ and number of Kondo channels $N$. In tetragonal systems this can happen for both 
easy-plane or easy axis anisotropy. The characteristic crossing of magnetic susceptibilities is not an artefact of the weak-coupling RG 
treatment but can be reproduced in brute-force perturbation theory. Employing numerical renormalization group (NRG), we show that for an
under-screened moment ($S=1$, $N=1$) with easy-plane anisotropy,  a crossing of magnetic susceptibilities can also occur in the 
strong-coupling regime (below the Kondo temperature).  This suggests that collective magnetic ordering of such under-screened moments 
would develop along the magnetic hard axis.  
\end{abstract}

\maketitle

\section{Introduction}

Fluctuations are at the heart of many complex ordering phenomena, leading to the formation of exotic phases of matter. 
Examples include nematic order in iron-based superconductors \cite{Cruz+08,Zhao+08}, driven by strong spin fluctuations above 
the magnetic ordering temperature \cite{Fernandez+14}, and $p$-wave spin-triplet superconductivity near 
ferromagnetic quantum critical points  \cite{Anderson+73,Fay+80}. In the latter case, the required attraction in the $p$-wave channel is generated by fluctuations.
This mechanism is  very similar in spirit to fluctuation generated Casimir and Van-der-Waals forces \cite{Green+18}.

In itinerant ferromagnets the coupling between the magnetic order parameter and soft electronic particle-hole fluctuations leads to a 
plethora of exotic ordering phenomena \cite{Brando+16}. It is responsible for fluctuation induced first-order behaviour at low temperatures 
\cite{Belitz+99,Chubukov+04,Belitz+05,Kirkpatrick+12},  observed experimentally in many systems 
\cite{Pfleiderer+01,Uemura+07,Otero+08,Taufourr+10,Yelland+11}.  Since the phase space for electronic fluctuations can be enhanced by deformations
of the Fermi surface, metallic ferromagnets are very susceptible towards the formation of spin nematic \cite{Chubukov+04}, modulated 
superconducting \cite{Conduit+13} or incommensurate magnetic order \cite{Conduit+09,Karahasanovic+12,Pedder+13,Taufour+16,Friedmann+18}. 

Fluctuations can also have counter-intuitive effects upon the direction of magnetic order parameters. A notable 
example is the partially ordered phase of  MnSi, in which the helimagnetic ordering vector rotates away from the lattice favored 
directions \cite{Pfleiderer+04,Kruger+12}. Magnetic hard-axis ordering in metallic ferromagnets is fairly wide spread \cite{Hafner+19,list}.
Such a moment re-orientation can arise as combined effect of fluctuations and magnetic frustration in a local moment model \cite{Andrade+14}. 
In an alternative scenario the effect was attributed to soft electronic particle-hole fluctuations in a purely itinerant model with spin-orbit induced 
anisotropy \cite{Kruger+14}.

In this Article we show that electronic fluctuations can drive magnetic hard-axis ordering
in anisotropic Kondo materials. As first established by Kondo \cite{Kondo64},  the scattering of electrons by local moments gives rise to 
logarithmic corrections to the magnetic susceptibility. In the presence of magnetic anisotropy, these logarithmic corrections depend upon direction. Near the
Kondo scale, these terms can completely overwhelm the crystal-field anisotropy experienced by the local moment, driving a moment re-orientation.  

We identify  a generic mechanism for magnetic hard-axis ordering that fully accounts for the following experimental facts \cite{Hafner+19}: 
(i) All the materials that show hard-axis ordering are Kondo systems.
(ii) The susceptibility crossing occurs above the magnetic ordering temperature $T_c$.
(iii) In tetragonal  systems the moment reorientation can occur from easy plane to hard axis \cite{Myers+99,Krellner+08,Jesche+12,Lausberg+13}, 
or the other way round, from easy axis to hard plane \cite{Steppke+13,Rai+19}. 
(iv) The effect also occurs in systems that show a first-order magnetic transition \cite{Muro+98,Anand+18}. 
(v) Similar magnetic hard-axis ordering is observed in Kondo systems that order antiferromagnetically \cite{Kondo+13,Khalyavin+13,Takeuchi+01}.

The paper is organized as follows. In Sec.~\ref{sec.model}, we introduce the $S\ge 1$, $N$-channel single-impurity Kondo model with single-ion anisotropy. The interplay of 
Kondo screening and anisotropy is studied within perturbative RG in Sec.~\ref{sec.pertRG}. We show that near the Kondo scale the single-ion anisotropy can change
sign, indicative of a reorientation of the dressed magnetic moment. As illustrated in Sec.~\ref{sec.resonance}, this reorientation might be interpreted as a resonance 
effect. In Sec.~\ref{sec.pert} we show that the effect can be reproduced by brute-force second-order perturbation theory, which shows a crossing of the magnetic 
susceptibilities for a large range of parameters and different values of $S$ and $N$. In Sec.~\ref{sec.NRG} we use the numerical renormalization group (NRG) to investigate the 
strong-coupling behavior of the single channel Kondo model of an $S=1$ spin with single-ion anisotropy and demonstrate that a crossing of the magnetic susceptibilities 
 can occur far below the Kondo temperature in systems with easy-plane anisotropy. Finally, in Sec.~\ref{sec.disc} we summarize and discuss our results.

\section{Kondo Model}
\label{sec.model}

Since the magnetic susceptibility crossing occurs above $T_c$, irrespective of the order of the transition and the nature of the 
 ordered state, the  dominant effect can be understood on the level of a single-impurity Kondo model 
with single-ion anisotropy,
\begin{eqnarray}
\label{eq.ham}
\hat{H} &=& \sum_{n=1}^N \sum_\bk^{|\epsilon_\bk|<\Lambda} \epsilon_\bk \boldsymbol{\psi}^{\dagger}_{\bk n} \boldsymbol{\psi}_{\bk n} + \alpha \Lambda \left(\hat{S}^z\right)^2\\
& & +  \frac{1}{N^2}\sum_{n=1}^N \sum_{\bk,\bq} \sum_{\gamma=x,y,z} J_\gamma \hat{S}^\gamma  \boldsymbol{\psi}^{\dagger}_{\bk n} \boldsymbol{\sigma}_\gamma \boldsymbol{\psi}_{\bq n}. 
\nonumber
\end{eqnarray}

Here $\boldsymbol{\psi}^{\dagger}_{\bk n} = (c^\dagger_{\bk n\uparrow}, c^\dagger_{\bk n\downarrow})$ with $c^\dagger_{\bk n\sigma}$ the creation operator of an electronic
quasiparticle with momentum $\bk$ and spin $\sigma$ in channel $n=1,\ldots N$. The first term simply denotes $N$ identical bands with dispersion $\epsilon_\bk$, subject to 
an energy cut-off $\Lambda$.   The second term is the single-ion anisotropy of the local moment spin ($S\ge 1$) in a tetragonal crystal, expressed in units of $\Lambda$. In the following we will investigate both easy-axis ($\alpha<0$) and easy-plane ($\alpha>0$) anisotropies. The last term in the Hamiltonian denotes the Kondo coupling between 
the impurity spin and the conduction electrons, where $\boldsymbol{\sigma}_\gamma$ are the standard Pauli matrices. Assuming tetragonal symmetry we have Kondo couplings $J_{xy}:=J_x=J_y$ and $J_z$.

\section{Perturbative RG} 
\label{sec.pertRG}

To analyze the scale dependence of  the single-ion anisotropy $\alpha$ and the Kondo couplings $J_{xy}$ and $J_z$, we integrate 
out processes to second order in the Kondo couplings that involve the creation of particles or holes in the infinitesimal energy shells $\Lambda e^{-d\ell} < |\epsilon_\bk| < \Lambda$.
This procedure, dubbed ``poor man's scaling'',  was first applied by Anderson \cite{Anderson70} to the anisotropic $S=1/2$ one-channel  
Kondo model. For $S=1/2$, anisotropy only enters through the Kondo couplings  $J_{xy}$ and $J_z$.

Here we generalize to $N$ channels and an $S\ge 1$ impurity subject to single-ion anisotropy. Moreover, in the spirit of the conventional RG
treatment we rescale to the original cut-off at each RG step. A detailed derivation of the general RG equations is given in Appendix \ref{ap.pertRG}. Since the qualitative 
behaviour of the RG flow is the same for all values of $S\ge 1$ and $N$, we focus on the under-screened case with $S=1$ and $N=1$ from now on. 
The weak-coupling RG equations are 
\begin{eqnarray}
\label{eq.RG}
\frac{d g_z}{d\ell}&=&g_{xy}^2(1+\alpha),\nn \\
\frac{d g_{xy}}{d\ell} & = &  g_{xy}g_z (1-\frac{\alpha}{2}),\\
\frac{d \alpha}{d\ell} &=& \alpha + g_{xy}^2- g_z^2   - 3 g_{xy}^2 \alpha,\nn
\end{eqnarray}
where $g_\gamma=2\rho J_\gamma$ are the dimensionless Kondo couplings. For simplicity, we have adopted the usual assumption \cite{Anderson70} of 
a constant density of states $\rho$. The scale parameter $\ell$ is related to temperature, $\ell = \log(\Lambda/T)$.

 In the absence of anisotropy, $g_{xy}=g_z$ and $\alpha=0$, the system remains isotropic 
under the RG flow, as expected. For $\alpha=0$, the RG equations for $g_{xy}$ and $g_z$ take the familiar form \cite{Anderson70}.
In the relevant regime of antiferromagnetic Kondo couplings the flow is towards strong coupling, $g_\gamma\to\infty$, corresponding to the Kondo regime. 

\begin{figure}
\centering
\includegraphics[width=\columnwidth]{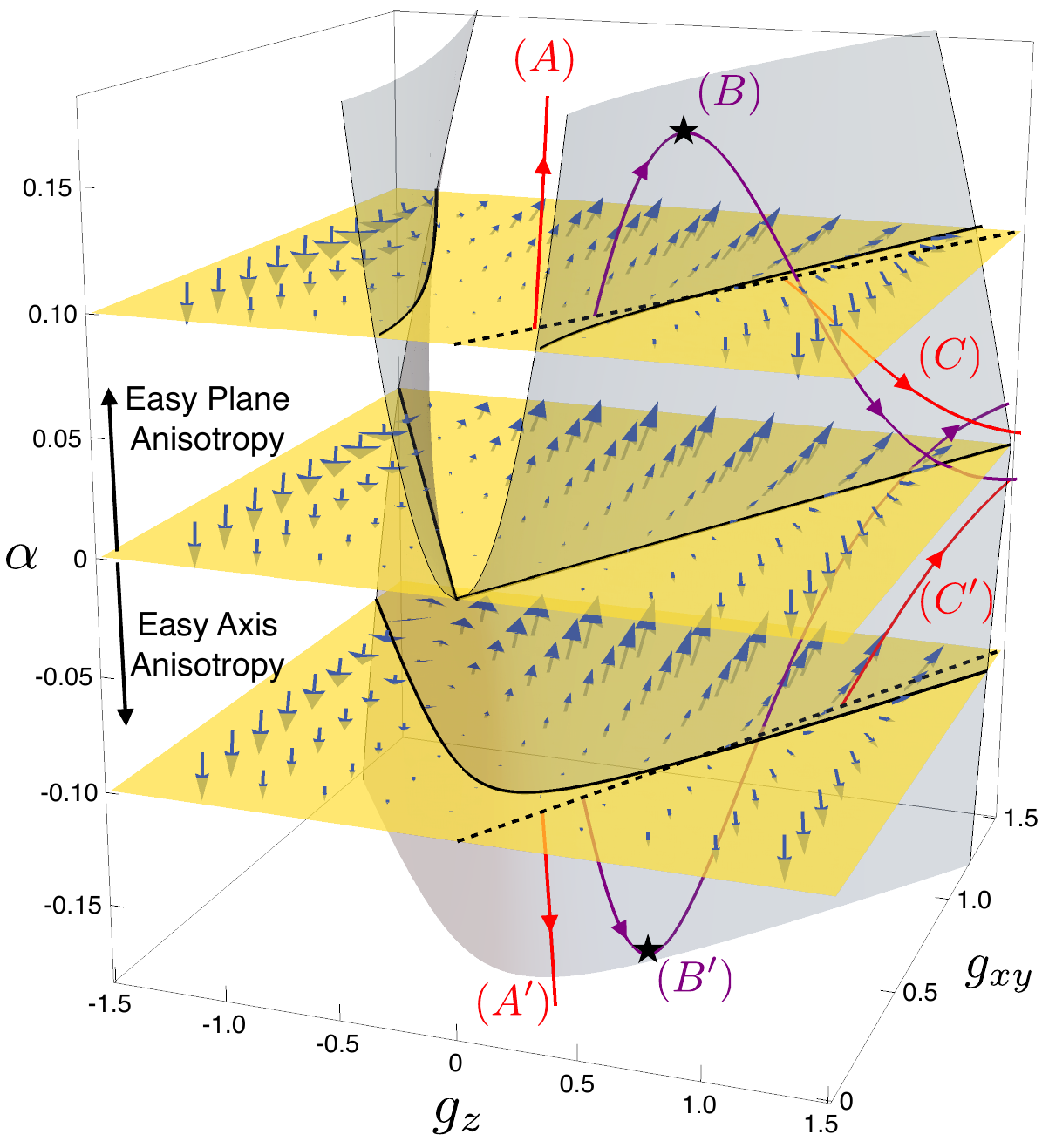}
\caption{RG flow of the single-ion anisotropy $\alpha$ and Kondo couplings $g_{xy}$, $g_z$. The direction of the flow  is shown 
by blue arrows and the transparent grey surface separates regions of increasing and decreasing $\alpha$. The trajectories $(A)$,$(B)$,$(C)$ correspond to an initial easy-plane anisotropy $\alpha(0)=0.1$ and increasing values of $g_{xy}(0)=g_z(0)$: 
$(A)$ For weak $g_\gamma(0)$ the anisotropy stabilizes the moment and supresses Kondo screening. $(B)$ For intermediate values of $g_\gamma(0)$, $\alpha$ 
changes sign before $g_\gamma$ diverge. This indicates
 a moment re-orientation above the Kondo temperature. $(C)$ For large $g_\gamma(0)$ the Kondo scale is reached before moment 
 re-orientation can occur. $(A')$,$(B')$ and $(C')$ show the analogous behaviour for an initial easy-axis anisotropy $\alpha(0)=-0.1$.}
\label{figure1}
\end{figure}

This picture is incomplete, however, since the anisotropy in the Kondo couplings generates 
single-ion anisotropy due to the $\left(g_{xy}^2- g_z^2 \right)$ term in the RG equation for $\alpha$. This leads to a flow out of the $\alpha=0$ plane. For $g_{xy}>g_z$
the flow is to positive $\alpha$, corresponding to easy-plane anisotropy, while for $g_{xy}<g_z$ an easy axis anisotropy is generated. Finite $\alpha$ modifies 
the RG flow of the Kondo couplings, e.g., easy-plane anisotropy ($\alpha>0$) leads to $g_z$ growing faster than $g_{xy}$. The interplay of these effects  initially leads to 
a ``restoration of symmetry" \cite{Konik+02} and ultimately to the moment re-orientation that is the main subject of this work.

In Fig.~\ref{figure1}, the evolution of the coupling constants under the RG is shown. 
For the trajectories $(A)$, $(B)$ and $(C)$  we have chosen an easy-plane anisotropy $\alpha(0)=0.1$ and initially isotropic Kondo couplings,  $g_{xy}(0)=g_z(0)$.
In the regime of weak Kondo coupling $(A)$, $\alpha$ keeps growing, leaving the regime where the RG equations are valid.  
This behaviour indicates that the single ion anisotropy stabilizes the moment, preventing Kondo screening. Magnetic hard-axis ordering therefore does not occur for sufficiently 
strong anisotropy, compared to the Kondo coupling, which is consistent with experimental observations \cite{Hafner+19}.

For $(B)$ the growing splitting of the increasing Kondo couplings reverse the flow of $\alpha$ at a scale $\ell_\textrm{max}$, 
corresponding to the point where the trajectory crosses the grey surface in Fig.~(\ref{figure1}), defined by $\alpha + g_{xy}^2- g_z^2   - 3 g_{xy}^2 \alpha=0$. At some scale 
$\ell_0$, corresponding to a temperature 
$T_0 = \Lambda e^{-\ell_0}$, $\alpha$ changes sign, indicating a re-orientation of the moment. At a larger scale $\ell_*>\ell_0$  the rapidly increasing Kondo couplings 
diverge, corresponding to the Kondo temperature $T_K=\Lambda e^{-\ell^*}<T_0$. The evolution of $g_{xy}(\ell)$, $g_{z}(\ell)$ and $\alpha(\ell)$ resulting in the trajectory $(B)$ are 
shown in Fig.~\ref{figure6} in Appendix \ref{ap.pertRG}.

If the initial Kondo couplings are too large $(C)$, the Kondo scale is reached before a moment re-orientation occurs.
Note that this strong coupling regime lies beyond the validity of the perturbative RG treatment. The trajectories 
$(A')$, $(B')$ and $(C')$ show the completely analogous behaviour for the case of easy-axis anisotropy.

\section{Reorientation as a resonance effect} 
\label{sec.resonance}

Treating the exchange between the impurity and the conduction electrons perturbatively, we can think of the local moment as being dressed with particle-hole fluctuations. If we trace out the conduction electrons with respect to the Gibbs thermal ensemble, we obtain the renormalized energies of the dressed $S=1$ impurity with quantum number $m_S=m$, 
\begin{eqnarray}
E(m)= \alpha \Lambda m^2 + 2 J^2 \sum_{\mathbf{k}, \mathbf{q},l}  \frac{n(\epsilon_{\mathbf{q}}) (1-n(\epsilon_{\mathbf{k}})) }{\alpha \Lambda (m^2-l^2) -( \epsilon_{\mathbf{k}} - \epsilon_{\mathbf{q}})},\quad
\end{eqnarray}
 where the $l$-sum is constrained by $1\le|m+l|\le 2$,  $J_{\gamma}=J$ \color{black} and $n(\epsilon )=1/[1+\exp(\epsilon/T)]$ is the Fermi function. In the limits $\alpha \ll 1$ and $\Lambda \gg T$, this expression evaluates to 
\begin{eqnarray}
E(m) = \alpha \Lambda m^2 \left( 1 - \frac{3}{2} g^2 \log \frac{\Lambda}{2T}\right) + {\rm const}.
\label{eq:energy}
\end{eqnarray}
We can see that Kondo screening leads to a reduction in the renormalised anisotropy $\alpha$ that grows with temperature, and can lead to change in sign of the renormalised $\alpha$ at low enough temperatures.  This is because of stronger resonance between impurity states with higher energy $\alpha\Lambda m^2$  and particle-hole excitations. 

\begin{figure}
\centering
\includegraphics[width=0.85\columnwidth]{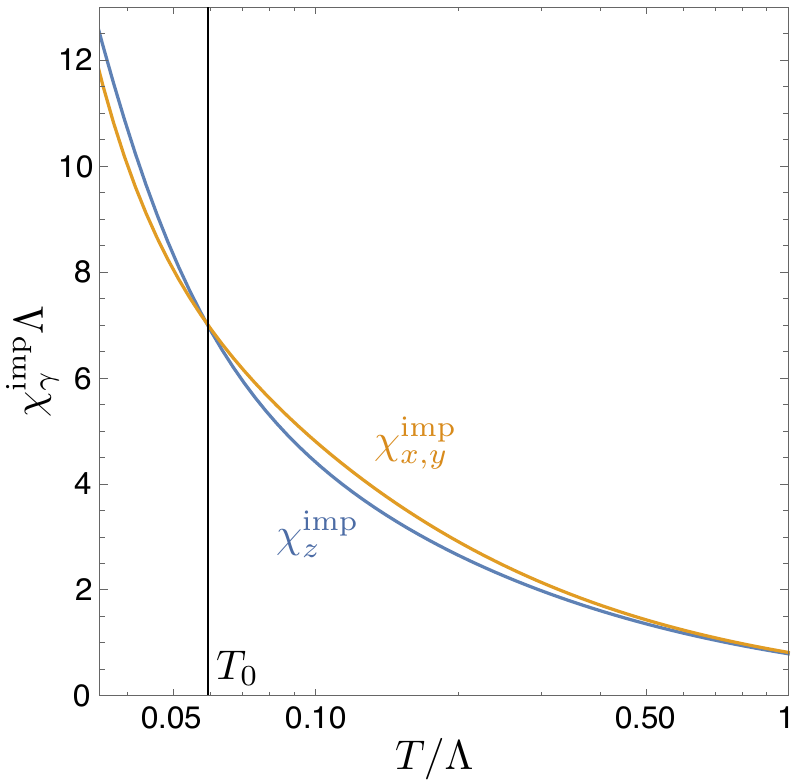}
\caption{Susceptibility crossings of the Kondo model with $S=1$, $N g_\gamma^2=0.23$ and easy-plane anisotropy $\alpha=0.05$ at $T_0>T_K$,  obtained by second-order perturbation theory.  
Such moment re-orientation can be found for all values of $N$ and $S\ge 1$ and both signs of $\alpha$.}
\label{figure2}
\end{figure}

\section{Impurity Susceptibility from perturbation theory} 
\label{sec.pert}

We have identified the moment re-orientation to a magnetic hard direction 
from the sign change of the single-ion anisotropy $\alpha$ under the RG flow. 
In the perturbative regime,  this should correspond to a crossing of the local magnetic susceptibilities  $\chi^{\rm imp}_z$ and $\chi^{\rm imp}_x= \chi^{\rm imp}_y$, which are defined as
\begin{eqnarray}
\chi^{\rm imp}_{\gamma}\equiv \int_0^{\beta } d\tau \langle \hat{S}^{\gamma}(\tau) \hat{S}^{\gamma}(0)\rangle.
\end{eqnarray}

Since for certain parameters the moment re-orientation occurs above  $T_K$ one should be able to observe the effect by calculating the magnetic 
susceptibilities in second-order perturbation theory in the Kondo couplings,
\begin{equation}
\chi_\gamma^{\rm imp} (T) = \chi^{\rm  free}_\gamma(T)+\sum_{\gamma'} F_{\gamma\gamma'}(T)g_{\gamma'}^2 +\mathcal{O}(g_\gamma^4),
\end{equation}
where $\chi^{\rm  free}$ denotes the susceptibility of a free $S=1$ impurity with single-ion anisotropy $\alpha$, 
\begin{equation}
\chi^{\rm  free}_{z}  =  \frac{2 e^{-\alpha\Lambda/T}}{T Z^{\rm imp}},\quad \chi^{\rm free}_{x,y}  =  \frac{2\left(1-e^{-\alpha\Lambda/T}\right)}{\alpha\Lambda Z^{\rm imp}},
\end{equation}
with $Z^{\rm imp} = 1+2e^{-\alpha\Lambda/T}$. This calculation was first performed by Kondo for the isotropic system \cite{Kondo64}
and later generalized to study the effects of a hexagonal crystal field in dilute alloys \cite{Borchi+74}. Unfortunately, the authors only calculated 
$\chi^{\rm imp}_z$ along the easy direction, not anticipating  a susceptibility crossing close to $T_K$ \cite{pert}. 

Since, to the best of our knowledge, results for $\chi^{\rm imp}_{x,y}$ are not available in the literature, we present a calculation 
in Appendix \ref{ap.2ndorder}, deriving explicit but lengthy expressions for $F_{xz}=F_{yz}$, $F_{xx}=F_{xy}=F_{yy}$ and $F_{zz}$. 
As anticipated, a crossing of the susceptibilities can be obtained 
in perturbation theory, regardless of the number of channels $N$, which simply enters in the prefactor $N g_\gamma^2$ of the perturbative 
correction,  for all  $S\ge 1$ and both signs of $\alpha$. In Fig.~\ref{figure2} we show an example of a susceptibility for $S=1$.

\section{Numerical Renormalization Group} 
\label{sec.NRG}

In order to investigate if hard-direction ordering of under-screened moments could occur in the strong-coupling regime 
at temperatures far below the Kondo temperature $T_K$, we employ the numerical renormalization group (NRG). Previous NRG studies \cite{Zitko+08} analyzed the effects 
of single-ion anisotropy on the Kondo screening mechanism and on possible non-Fermi-liquid behaviour, but did not investigate the behaviour of magnetic susceptibilities along 
different directions.  The calculation of transverse magnetic susceptibilities using NRG has been reported in the literature for related models \cite{Cano+13}.
It is important to stress that in the strong-coupling regime the physics will crucially depend on $S$ and $N$. Here we only investigate the single-channel 
Kondo model for $S=1$. Details on our NRG calculations can be found in Appendix \ref{ap.NRG}.

In the strong-coupling regime, the experimentally relevant quantity  is not $\chi_\gamma^{\rm imp}$, but the impurity contribution to the total susceptibility, $\chi_\gamma^{\rm cont}$, 
defined as the difference between the total susceptibility of the system with and without the impurity. As temperature is lowered, the impurity increasingly 'outsources' its 
magnetic moment to the conduction electrons. While the total $z$-angular momentum is conserved and the dressed impurity states are eigenstates of $\hat{J}_z $, 
the conduction electrons are carrying an increasing fraction of the impurity's angular momentum which is no longer negligible at $T\sim T_K$. 

\begin{figure}
\centering
\includegraphics[width=0.85\columnwidth]{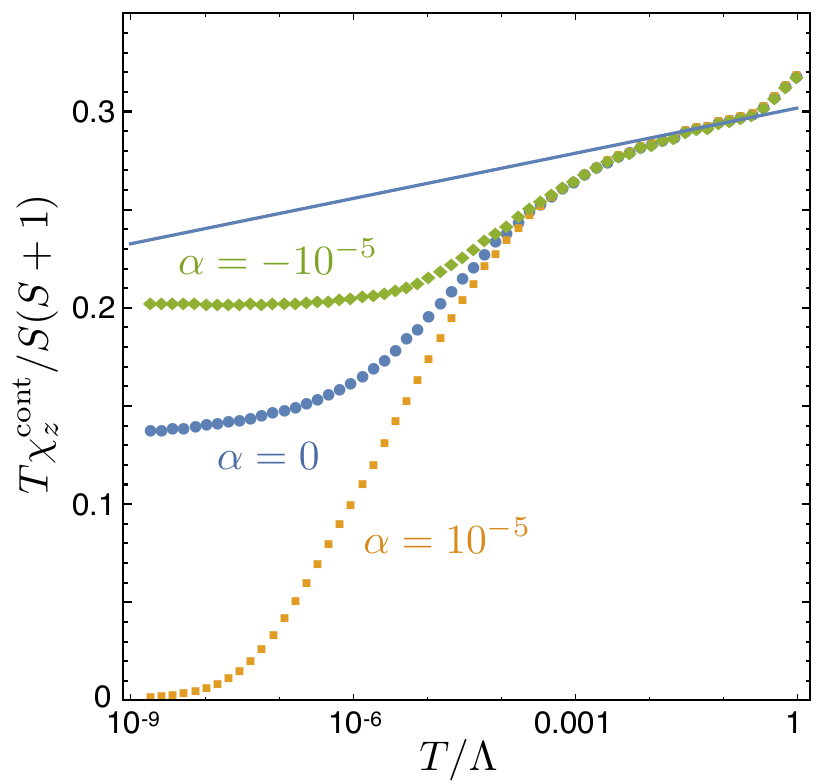}
\caption{Impurity contribution to the susceptibility $T\chi_z^{\rm cont}/S(S+1)$ for single-channel 
$S=1$ Kondo models, computed by NRG. The Kondo coupling $g_\gamma=0.1$ results in a Kondo temperature of $T_K \sim 10^{-5} \Lambda$. The different curves show the 
isotropic case ($\alpha=0$), and the results for easy-plane ($\alpha=10^{-5}$) and easy-axis  ($\alpha = -10^{-5}$) anisotropies.  The results presented here show excellent agreement 
with the NRG results of Ref.~\cite{Zitko+08}. The line shows the result from second order perturbation theory for the isotropic case.}
\label{figure3}
\end{figure}

We first benchmark our NRG results against those of Ref.~\cite{Zitko+08}, where the total susceptibility $\chi^{\rm cont}_z$ in the $z$-direction was calculated for systems with easy-plane and 
easy-axis anisotropy. Our NRG results, presented in Fig.~\ref{figure3} show excellent agreement with the results of that work. However, the reference did not include the total susceptibility in the 
$x$-direction $\chi^{\rm cont}_x$, which is a dynamical, rather than thermodynamic quantity, as $[\hat{H},\hat{J}_x] \neq 0$. In this case, the computation of $\chi^{\rm cont}_x$ is equivalent 
to calculating the entire spectral density function, which is a more involved process \cite{SM}.

\begin{figure}
\centering
\includegraphics[width=\columnwidth]{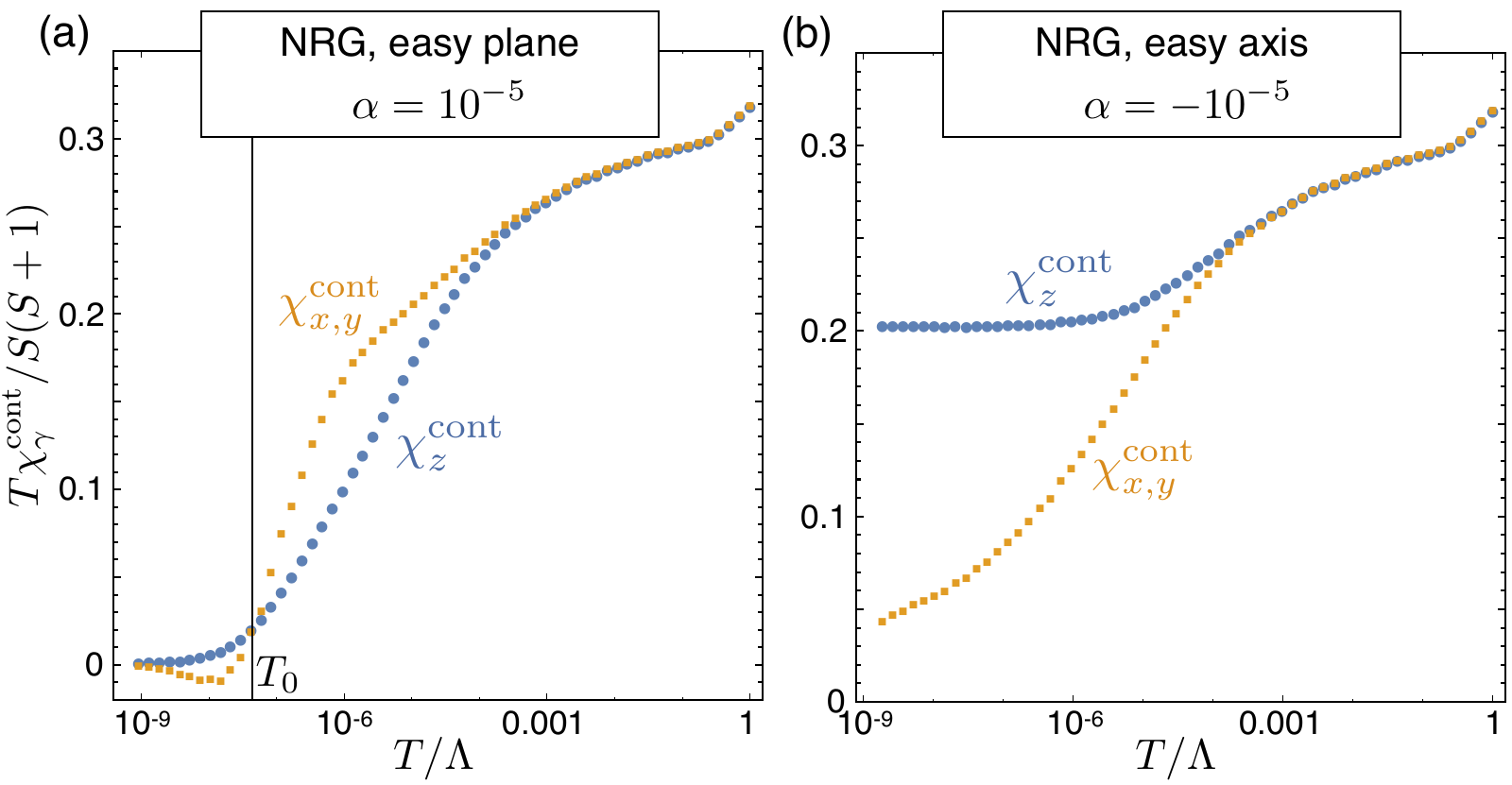}
\caption{Impurity contribution to the susceptibility for the single-channel $S=1$ Kondo models with $g_\gamma=0.1$ and (a) easy-plane ($\alpha=10^{-5}$) and (b) easy-axis ($\alpha = -10^{-5}$) anisotropies. Note that moment re-orientation, signalled by a crossing of the susceptibilities, only occurs in the case of easy-plane anisotropy.}
\label{figure4}
\end{figure}

Fig.~\ref{figure4} shows the total susceptibilities $\chi_\gamma^{\rm cont}$ for the same parameters  of the anisotropic Kondo model ($S=1$, $N=1$, $g_\gamma=0.1$, $\alpha=\pm 10^{-5}$)
along both, the $z$-axis and directions in the $xy$-plane. Unlike in the weak-coupling regime where moment re-orientation can occur regardless of the sign of $\alpha$, at strong coupling 
($T\ll T_K$) we only observe a crossing of magnetic susceptibilities in the case of easy-plane anisotropy ($\alpha>0$). 

The crossing of total susceptibilities can be understood in terms of the subspaces with different total angular momentum $J_z$. Without Kondo screening the states are product states of the impurity 
and conduction electrons. We can  divide the $J_z=0$ subspace into sectors $(m, n) = (0,0), (1,-1), (-1,1)$, where $m, n$ are the $z$-angular momenta of the impurity and conduction 
electrons, respectively. In the limit $T\ll \Lambda$, it costs energy to inject angular momentum into the Fermi sea and it costs energy for the impurity to have $m\neq 0$. Hence, sectors $(1,-1), (-1,1)$ 
are higher in energy than the $(0,0)$ sector and Kondo exchange gives weak mixing between the sectors. For the $J_z=1$ subspace the relevant sectors are $(m,n) = (1,0), (0,1)$. These 
are much closer in energy than $J_z=0$ sectors, because the cost of injecting angular momentum into the Fermi sea can be offset by lowering the $m$ quantum number of the impurity. The sectors 
resonate more strongly  and Kondo exchange gives stronger mixing between them. As a result, the $J_z=1$ subspace is lowered in energy more strongly than the $J_z=0$ subspace, allowing 
for the possibility of a susceptibility crossing at $T\ll T_K$.

The negative susceptibility contributions at lowest temperatures in Fig.~\ref{figure4}(a) result from the discretization of the conduction electron band in Wilson's NRG \cite{Zhuravlev09,Fang+15}. 
This is equivalent to being away from the thermodynamic limit, where there is a finite number of conduction electron sites $N_s$, resulting in a non-zero Curie moment  
$T\chi$ of the free conduction electrons. When the impurity becomes entangled with the conduction electrons, it leads to a reduction of the conduction electron's Curie moment.  
This can lead to a negative contribution contribution in the difference of total susceptibilities with and without the impurity. The susceptibility crossing 
 takes place above the temperature where the contribution becomes negative and is robust against changes of the discretization. To some extent the numerical discretization mimics that 
 in a realistic system there is a finite density of impurities and hence  a finite number of conduction electron sites per impurity, even in the thermodynamic limit. It would be interesting to self-consistently 
 account for the Kondo-lattice using dynamical mean-field theory.

\section{Reorientation in the infinitely narrow bandwidth limit} 

For the sake of completeness, we note here that we also observe magnetic reorientation in the less experimentally relevant limit of an infinitely narrow band ($\Lambda \rightarrow 0$), where it suffices to look at a single electron site coupled to the impurity. In the case $S=1$ this results in the Hamiltonian
\begin{equation}
	\hat{H}=g[\hat{S}^{+}c^{+}_{\downarrow}c_{\uparrow}+\hat{S}^{-}c^{+}_{\uparrow}c_{\downarrow}+\hat{S^z} (c^+_{\uparrow}c_{\uparrow}-c^+_{\downarrow}c_{\downarrow})]+\alpha (\hat{S^z})^2,
\end{equation}
where for simplicity we keep the Kondo coupling $g=g_\gamma$ to be isotropic. The Hilbert space contains a total of 12 states, and since the Hamiltonian conserves the total spin, the $12\times12$ matrix 
is block-diagonal and the largest matrices to be diagonalized are two $2\times2$ matrices which can be done by hand. From the eigenstates $|i\rangle$ and corresponding energies $E_i$, it is straightforward 
to compute the total susceptibilities along different directions,
\begin{eqnarray}
\chi_z^\textrm{tot} & = & \frac{1}{T} \sum_{i}  e^{-E_{i}  / T}\langle  i| (\hat{J}^z)^2 | i \rangle,\\
\chi_x^\textrm{tot} & = &  \sum_{i, j} \frac{e^{-E_{i}/T}- e^{-E_{j}/T} }{E_{j} - E_{i}} \left| \langle i |    \hat{J}^x  
|j \rangle \right|^2,
\end{eqnarray}
where $\hat{J}^\alpha = \hat{S}^\alpha+\hat{s}^\alpha$ denote the total spin operators. 

\begin{figure}[t!]
\centering
\includegraphics[width=\columnwidth]{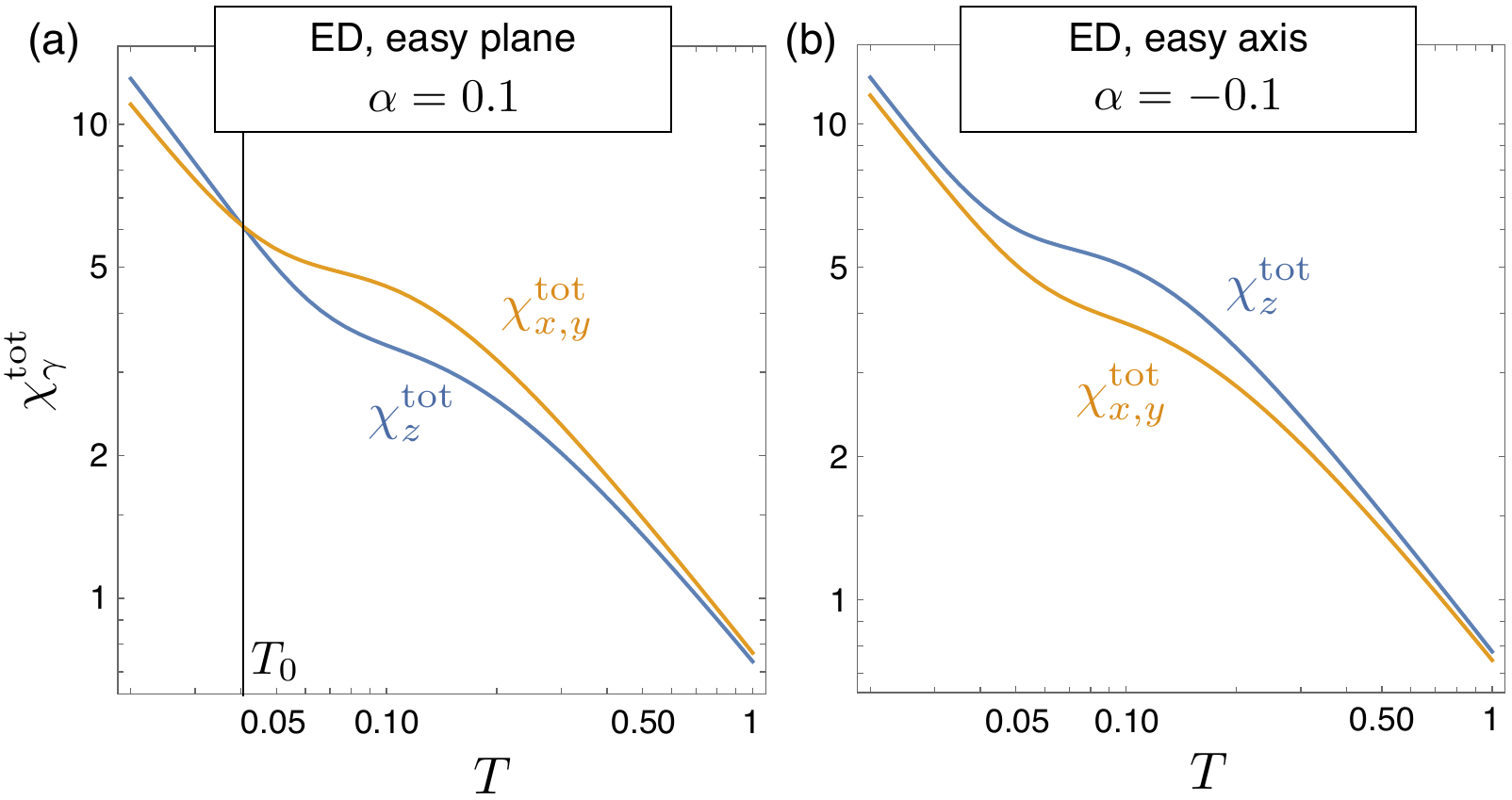}
\caption{The total susceptibilities $\chi^\textrm{tot}_z$ and $\chi^{tot}_{x,y}$ in the infinitely narrow bandwidth limit, for (a) $S=1, g=0.1, \alpha=0.1$, and (b) $S=1, g=0.1, \alpha=-0.1$. Moment reorientation 
happens in the case of easy-plane anisotropy.}
\label{figure5}
\end{figure}

The total susceptibilities are shown in Fig.~\ref{figure5}. Similar to the NRG results, for $S=1$ there is only a crossing of total susceptibilities for $\alpha>0$, the easy-plane case, but not the other way around. 
We would like to stress again that although the NRG results are similar, they are valid in a very different limit.

\section{Discussion} 
\label{sec.disc}

We have presented a perturbative RG analysis of the single-impurity Kondo model with single-ion anisotropy. Our main 
finding is that fluctuations near the Kondo temperature $T_K$ can drive a re-orientation of the moment away from the lattice favoured direction at $T_0>T_K$. 
This hard-direction ordering occurs over a wide range of parameters and for different types of anisotropy. As additional proof of principle, we have shown that a 
crossing of magnetic susceptibilities occurs in second-order perturbation theory, for all values of $N$ and $S\ge 1$. 

It is important to stress that magnetic hard-direction ordering could occur even at temperatures above $T_0$, since  the RKKY interaction,  
$J_\gamma^{\rm RKKY}\sim g_\gamma^2$, is significantly enhanced along the hard direction, e.g. for a system with easy-plane anisotropy we find $g_z > g_{xy}$ 
significantly above $T_0$. As a result, the susceptibility along this direction would diverge first, giving rise to magnetic hard-direction ordering.
 
Using NRG, we investigated the strong-coupling behaviour of the under-screened $S=1$ single channel Kondo model with single-ion anisotropy.
We found that in this regime a crossing of magnetic susceptibilities can occur, but only in systems with easy-plane anisotropy. While the NRG results are robust at strong coupling, 
they crucially depend on the Hilbert space truncation and energy discretization in the intermediate temperature regime \cite{Zitko11,Yang+20}, making it impossible to numerically resolve susceptibility 
crossings and compare with our perturbative calculations. One would expect that with increasing coupling strength the effect becomes more asymmetric and eventually only survives in systems with 
easy-plane anisotropy.   This might explain why this case is more frequently 
observed in experiments \cite{Hafner+19}. It would be interesting to investigate the strong coupling behaviour of Kondo models with different $S$ and $N$, as well as  of  
 closely related Coqblin-Schrieffer models, which better describe systems with strong spin-orbit coupling. For the latter, we found susceptibility crossings in 
perturbation theory and at strong coupling in the infinitely narrow band limit. 

The mechanism presented here is rooted in the interplay of Kondo 
fluctuations and anisotropy on the single-impurity level. This would explain why hard-direction ordering is observed  in a range of Kondo lattice systems, irrespective of 
the order and universality of the magnetic phase transition \cite{Hafner+19} and for both ferromagnetic \cite{list} and antiferromagnetic 
ordering \cite{Kondo+13,Khalyavin+13,Takeuchi+01}. 

Advances in Nanotechnology and scanning tunnelling microscopy have led to a revival of the Kondo effect \cite{Kouwenhoven+01}, thanks to unprecedented control on 
the level of single magnetic ad-atoms on metallic surfaces \cite{Li+98,Madhavan+98,Knorr+02} or artificial magnetic elements in quantum dots \cite{Goldhaber+98,Potok+07}. 
Such experiments could in principle probe the fluctuation-driven re-orientation of a single magnetic impurity.

 We argue that the magnetic hard-direction ordering observed in a wide range of Kondo materials is predominantly driven by strong Kondo fluctuations. 
This mechanism might be further enhanced by soft electronic particle-hole fluctuations that can lead to moment re-orientation 
near ferromagnetic critical points \cite{Kruger+14}. Such a combined mechanism could be at play in YbNi$_4$P$_2$ which shows strong 
quantum critical fluctuations  \cite{Steppke+13}. 

Our work shows that strong fluctuations in anisotropic Kondo materials can drive magnetic hard-direction ordering. It is to be expected that the interplay of collective 
critical fluctuations and Kondo physics will lead to many more unexpected ordering phenomena that are yet to be revealed.

{\bf{Acknowledgements}}
The authors benefitted from stimulating discussions with M. Brando, H.-U. Desgranges, D. Hafner, A. Huxley, and A. Nevidomskyy. This work has been supported by the EPSRC through 
grant EP/P013449/1. 

\appendix

\section{Derivation of the perturbative RG equations}
\label{ap.pertRG}

The partition function of the single-impurity Kondo model (\ref{eq.ham})  can be written down in the path-integral,
\begin{eqnarray}
\mathcal{Z} = \int \mathcal{D} \mathbf{S}(\tau)\; \mathcal{D} [\boldsymbol{\psi}^{\dagger}_{\mathbf{k} n} (\tau)
\boldsymbol{\psi}_{\mathbf{k} n} (\tau)]\;
 e^{- \mathcal{S}_0
-\mathcal{U}
 },\label{eq1}
\end{eqnarray}
where $\mathbf{S}(\tau)=(S^x(\tau),S^y(\tau),S^z(\tau))$ is the impurity spin, $\boldsymbol{\psi}^{\dagger}_{\mathbf{k}n}(\tau)=(c^{\dagger}_{\mathbf{k}n\uparrow}(\tau),c_{\mathbf{k}n\downarrow}(\tau))$ the Grassmann variables describing the conduction electrons, and
\begin{eqnarray}
\mathcal{S}_{0} &=&\int_0^{\beta} d\tau \sum_{n=1}^N \sum_\bk^{|\epsilon_\bk|<\Lambda} \boldsymbol{\psi}^{\dagger}_{\bk n} (\tau)
\left(\partial_{\tau} + \epsilon_{\bk}
\right)
 \boldsymbol{\psi}_{\bk n} (\tau) \\
 & & + \mathcal{S}_{\rm imp}[\mathbf{S}(\tau)]  ,
\nonumber\\
\mathcal{U} &=& \int_0^{\beta} d\tau  \sum_{n=1}^N \sum_{\mathbf{k}, \mathbf{q}} \sum_{\gamma=x,y,z} J_\gamma S^\gamma(\tau)  \boldsymbol{\psi}^{\dagger}_{\bk n} (\tau) \boldsymbol{\sigma}_\gamma \boldsymbol{\psi}_{\bq n} (\tau).
\nonumber\\
\end{eqnarray}
$\mathcal{S}_0$ is the action corresponding to the Hamiltonian $\hat{H}_0=\sum_{\bk ,n,\sigma}\epsilon_{\bk} c^{\dagger}_{\bk n\sigma}c_{\bk n\sigma} + \alpha\Lambda ( \hat{S}^z )^2$ and $\mathcal{S}_{\rm imp}[\mathbf{S}(\tau)] $ is the action of a free impurity with a spin quantum number $S$ and Hamiltonian $\hat{H}_{\rm imp} = \alpha\Lambda ( \hat{S}^z )^2$. This latter action  contains the necessary terms that enforce constraints satisfied by $\mathbf{S}(\tau)$. We do not give an explicit form here, which will depend on the representation used, e.g. Abrikosov pseudofermions, spin-coherent states, Schwinger bosons, etc., and is not important for subsequent representation-independent calculations. Note also that every sum over electron momenta $\mathbf{k}$ includes the normalization factor of $1/\sqrt{N_s}$, where $N_s$ is the number of electron lattice sites.

\subsection{Perturbative Corrections}

We begin the renormalization group procedure by integrating out 'fast' fermion modes $\boldsymbol{\psi}^{\dagger}_{\bk n} (\tau)$ with $\Lambda e^{-d\ell}<|\epsilon_{\bk}|<\Lambda$. To second order in the Kondo exchange $\mathcal{U}$, the renormalized actions can be written as
\begin{eqnarray}
\mathcal{S}_{0}' &=& \int_0^{\beta} d\tau  \sum_{\bk, n}^{|\epsilon_\bk|<\Lambda e^{-d\ell}} \boldsymbol{\psi}^{\dagger}_{\bk n} (\tau)
\left(\partial_{\tau} + \epsilon_{\bk}
\right)
 \boldsymbol{\psi}_{\bk n} (\tau) \nn\\
 & & +\mathcal{S}_{\rm imp}[\mathbf{S}(\tau)]\\
 \mathcal{U} ' & =& \int_0^{\beta} d\tau   \sum_{\mathbf{k}, \mathbf{q}, n}^{|\epsilon_{\bk, \bq}|<\Lambda e^{-d\ell}} \sum_{\gamma=x,y,z} J_\gamma S^\gamma(\tau)  \boldsymbol{\psi}^{\dagger}_{\bk n} (\tau) \boldsymbol{\sigma}_\gamma \boldsymbol{\psi}_{\bq n} (\tau)\nn\\
& &  - \frac{1}{2}\langle \mathcal{U} ^2\rangle_{\rm fast}^{\rm conn.},
\end{eqnarray}
where the expectation value is taken with respect to the part of $\mathcal{S}_{0}$ describing the fast modes. We write $-\frac{1}{2}\langle \mathcal{U} ^2\rangle_{\rm fast}^{\rm conn.}$ as the following sum
\begin{eqnarray}
-\frac{1}{2}\langle \mathcal{U} ^2\rangle_{\rm fast}^{\rm conn.} = 
u_{1}J_{xy}^2 +u_2J_z^2   + u_3J_zJ_{xy} ,
\end{eqnarray}
and we will now go through the calculation of each of the coefficients $u_1,u_2,u_3$. The $J_{xy}^2 $ coefficient $u_1$ is given by the following expectation value

\begin{eqnarray}
u_1&=&-\int d\tau_1 d\tau_2 \sum_{n,m=1}^N\sum_{\bk,\bk'}^{<}\sum_{\bq,\bq'}^{>}\nn\\
& & \Big{(} S^+(\tau_2) S^-(\tau_1)\; c_{\bk m\uparrow}(\tau_2) c^{\dagger}_{\bk' n\uparrow}(\tau_1)\;\nn\\
& & \times\left\langle  c_{\bq m\downarrow}^{\dagger}(\tau_2) c_{\bq' n\downarrow} (\tau_1)\right\rangle
\nonumber\\
&&+ S^-(\tau_2)   S^+(\tau_1) \;
  c_{\bk m\downarrow} (\tau_2) c_{\bk' n\downarrow}^{\dagger}(\tau_1)\;\nn\\
  & & \times
\left\langle c^{\dagger}_{\bq m\uparrow}(\tau_2)  c_{\bq' n\uparrow}(\tau_1)\right \rangle
\Big{)}
\nonumber\\
&=&
-\rho d\ell\Lambda\int d\tau_1 d\tau_2 \sum_n\sum_{\bk,\bk'}^<
{\rm sign}(\tau_2 -\tau_1)e^{-|\tau_2 - \tau_1| \Lambda}\nn\\
& & \times\Big{(}
 S^+(\tau_2) S^-(\tau_1)\; c_{\bk n\uparrow}(\tau_2) c^{\dagger}_{\bk' n\uparrow}(\tau_1)\;
\nonumber\\
&&+ S^-(\tau_2)   S^+(\tau_1) \;
  c_{\bk n\downarrow} (\tau_2) c_{\bk' n\downarrow}^{\dagger}(\tau_1)
\Big{)}\nn\\
&   \equiv & u_{1}^{(1)}+u_{1}^{(2)}, \label{eq: u_1}
\end{eqnarray}
where we have used that $\sum_{\bq, \bq'}^>\langle  c_{\bq m\sigma}^{\dagger}(\tau_2) c_{\bq' n\sigma} (\tau_1)\rangle=\delta_{mn}\rho d\ell\Lambda {\rm sign}(\tau_2 -\tau_1)e^{-|\tau_2 - \tau_1| \Lambda}$ 
in the limit $\beta\Lambda\rightarrow\infty$. It is useful to introduce new variables $\tau = \frac{1}{2}(\tau_1+\tau_2)$ and $\Delta=\tau_2-\tau_1$, and split the integration over $\Delta$ into the 
$\Delta>0$ and $\Delta<0$ regimes. 
\begin{eqnarray}
u_{1}^{(1)} &= &-\rho d\ell \Lambda \sum_n \sum_{\bk,\bk'}^{<} 
\int d \tau \Big(\int_{\Delta>0} d \Delta \;e^{-\Lambda \Delta}\; \nn\\
& & \times\langle \tau |e^{\Delta \hat{H}_{0}/2}  \hat{c}_{\bk n\uparrow}\hat{S}^+e^{-\Delta \hat{H}_{0}}
 \hat{c}^{\dagger}_{\bk' n\uparrow} \hat{S}^- e^{\Delta \hat{H}_{0}/2}
 | \tau \rangle\nonumber\\
 &&  \;\;+\int_{\Delta<0} d \Delta \; e^{\Lambda \Delta}\;\nn\\
&& \times \langle \tau|e^{-\Delta \hat{H}_{0}/2}     \hat{c}^{\dagger}_{\bk' n\uparrow}  \hat{S}^- e^{\Delta \hat{H}_{0}}\hat{c}_{\bk n\uparrow}\hat{S}^+e^{-\Delta \hat{H}_{0}/2}|\tau \rangle \Big),\nn\\
\end{eqnarray}
where we have transformed to the operator representation of the expansion of the partition function in powers of $J_{\gamma}$, and $|\tau \rangle$ are the path-integral coherent states in terms of
 which the partition function is written down. Integrating over $\Delta$, we obtain
\begin{eqnarray}
 u_{1}^{(1)}&=&- \rho d\ell \Lambda  \sum_n\sum_{\bk,\bk'}^<\int d\tau \;\nn\\ 
& &  
 \left\langle \tau \left |
 \frac{\hat{S}^+\hat{S}^- \hat{c}_{\bk n\uparrow}  \hat{c}^{\dagger}_{\bk' n\uparrow}}{\Lambda(1+\alpha(1-2\hat{S}^z))}  
 +  \frac{\hat{S}^-\hat{S}^+ \hat{c}^{\dagger}_{\bk' n\uparrow}\hat{c}_{\bk n\uparrow}  }{\Lambda(1+\alpha(1+2\hat{S}^z))}  
\right |\tau \right \rangle 
\nonumber\\
&=&2 \rho d\ell  \sum_n\sum_{\bk,\bk'}^< 
\int d\tau \;
c^{\dagger}_{\bk n\uparrow}
 (\tau)c_{\bk' n\uparrow} (\tau)\nn\\
 & & \times\left(   \left(1 - \alpha \left(1-2S(S+1)\right) \right) S^z(\tau) -2\alpha \left(S^z(\tau)\right)^3 \right)
 \nonumber\\
&& -2 N\rho^2 \Lambda d\ell \int d\tau \;\left(1-\alpha(1-2S^z\left(\tau\right)\right)\nn\\
& & \times\left(S(S+1) - \left(S^z(\tau)\right)^2 +S^z(\tau)\right) + \mathcal{O} (\alpha^2),
\end{eqnarray}
where we have neglected terms proportional to $\epsilon_{\mathbf{k}}\Lambda^{-1}$ in the first line and terms second order in $\alpha$ in the second line. Following the same steps, we obtain for the second term in Eq.~\ref{eq: u_1},
\begin{eqnarray}
 u_{1}^{(2)} &=& - \rho d\ell \Lambda  \sum_n\sum_{\bk,\bk'}^< \int d\tau \;\nn\\
 & & \left\langle \tau \left|
 \frac{\hat{S}^-\hat{S}^+ \hat{c}_{\bk n\downarrow}  \hat{c}^{\dagger}_{\bk' n\downarrow}}{\Lambda(1+\alpha(1+2\hat{S}^z))}  
 +  \frac{\hat{S}^+\hat{S}^- \hat{c}^{\dagger}_{\bk' n\downarrow}\hat{c}_{\bk n\downarrow}  }{\Lambda(1+\alpha(1-2\hat{S}^z))}  
 \right|\tau \right\rangle 
\nonumber\\
&=&
-2 \rho d\ell \sum_n \sum_{\bk,\bk'}^<
 \int d\tau \;
  c^{\dagger}_{\bk n\downarrow}
 (\tau)c_{\bk' n\downarrow} (\tau)\nn\\
& & \times  \left(   \left(1 - \alpha \left(1-2S(S+1)\right) \right) S^z(\tau)
 -2\alpha \left(S^z(\tau)\right)^3 \right)
 \nonumber\\
 &&
 -2 N\rho^2 \Lambda d\ell 
 \int d\tau \; 
 \left(1-\alpha(1+2S^z(\tau)) \right)\nn\\
& & \times \left(S(S+1) - \left(S^z(\tau)\right)^2 -S^z(\tau)\right).
 \nonumber\\
\end{eqnarray}
Putting $u_{1}^{(1)}$ and $u_{1}^{(2)}$ together, we obtain the coefficient of the $J_{xy}^2$ term,
\begin{eqnarray}
u_1 &=&
2 \rho d\ell \sum_n \sum_{\bk,\bk'}^< 
\int d\tau \; \nn\\
& & \left(c^{\dagger}_{\bk n\uparrow} (\tau)c_{\bk' n\uparrow} (\tau) -c^{\dagger}_{\bk n\downarrow} (\tau)c_{\bk' n\downarrow} (\tau) \right)\nn\\
& & \times\left(   \left(1 - \alpha \left(1-2S(S+1)\right) \right) S^z(\tau) -2\alpha \left(S^z(\tau)\right)^3 \right)
 \nonumber\\
&& 
+4N\rho^2 d\ell (1-3\alpha) \Lambda
\int d\tau \;
\left(S^z(\tau)\right)^2 +\mathcal{O} (\alpha^2).
\end{eqnarray}

The $J_{xy}^2$ coefficient contains terms proportional to $\left(S^z(\tau)\right)^2$, which renormalize $\alpha$  by $4NJ_{xy}^2 \rho^2 d\ell (1-3\alpha)$, as well as terms 
proportional to $c^{\dagger}_{\sigma} (\tau)c_{\sigma} (\tau)S^z(\tau)$ and $c^{\dagger}_{\sigma}(\tau)c_{\sigma} (\tau)\left(S^z(\tau)\right)^3$.
For $S=1$, $\left(S^z(\tau)\right)^{3} = S^z(\tau)$, and both of these renormalize $J_{z}$.
For $S>1$, terms proportional to $c^{\dagger}_{\sigma}(\tau)c_{\sigma} (\tau)\left(S^z(\tau)\right)^3$ generate a new coupling that is not present in the original model. We will neglect these for 
$S>1$ and only keep terms proportional to $c^{\dagger}_{\sigma}
 (\tau)c_{\sigma} (\tau)S^z(\tau)$, which renormalize $J_z$. (Note that our result is exact for $S=1$.) We thus obtain the following renormalization of $J_z$

\begin{eqnarray}
\label{eq.rg1}
d J_{z} &=& 
2\rho  J_{xy}^2 d\ell \left\{
    \begin{array}{ll}
       (1+\alpha) & \mbox{for } S=1
       \nn\\ \nn\\
     \left[1 - \alpha \left(1-2S(S+1)\right) \right]  & \mbox{for }S>1
    \end{array} \right.\\
\end{eqnarray}

\begin{figure}
\centering
\includegraphics[width=0.95\columnwidth]{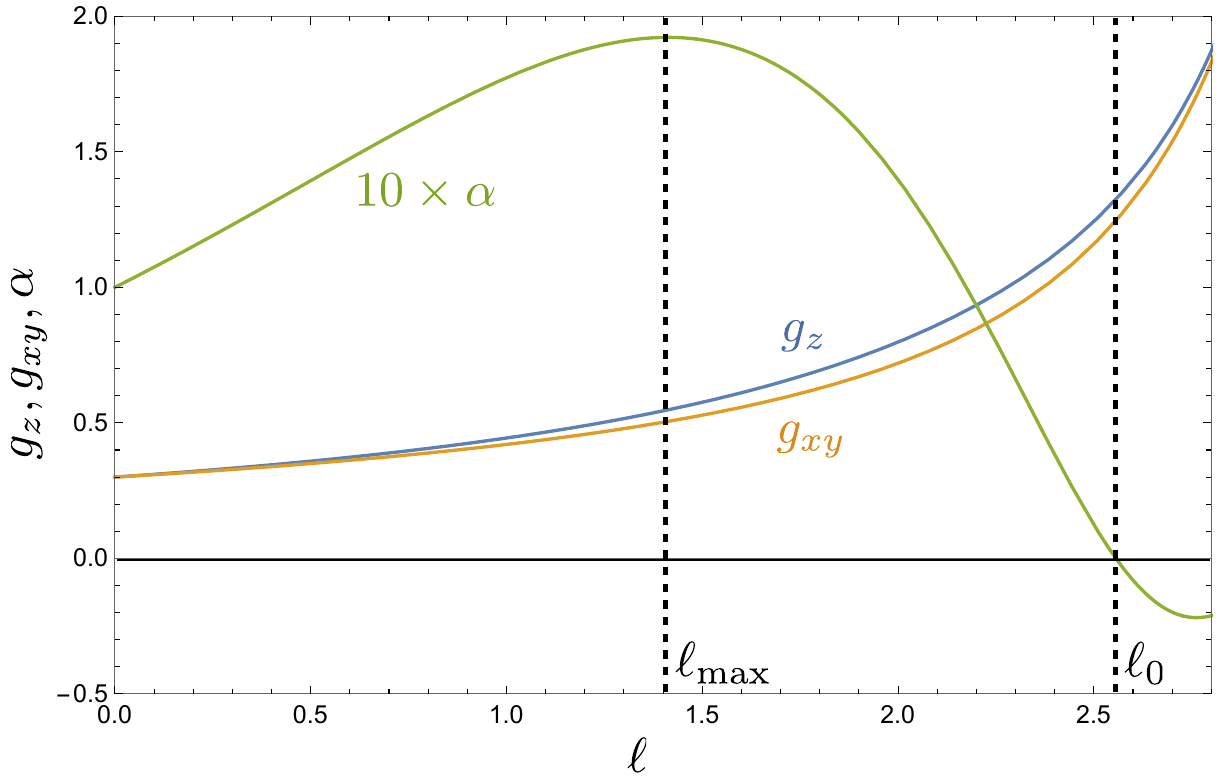}
\caption{The evolution of $\alpha$, $g_z$ and $g_{xy}$ as a function of $\ell$ for the parameters $S=1$, $N=1$, $\alpha(0)=0.1$, $g_\gamma(0)=0.3$,  corresponding to the trajectory (B) in Fig.~1.}
\label{figure6}
\end{figure}

\noindent
The $J_z^2$ coefficient $u_2$ is given by
\begin{eqnarray}
u_2 &=&
-\sum_{\sigma}\sum_{n,m=1}^N \sum_{\bk,\bk'}^< 
\sum_{\bq,\bq'}^> \int d\tau_1 d\tau_2 \;
 S^z (\tau_2)S^z(\tau_1) \; \nn\\
 & & c_{\bk m\sigma} (\tau_2)c^{\dagger}_{\bk' n\sigma}(\tau_1)
\; 
\left \langle c^{\dagger}_{\bq' m\sigma} (\tau_2) c_{\bq n\sigma}(\tau_1) \right\rangle \nn\\
& = &  -4N \rho^2 d\ell \int d\tau \;\left(S^z(\tau)\right)^2,
\end{eqnarray}
where we have followed the same steps as for the $J_{xy}^2$ coefficient $u_1$.
Hence, the $J_z^2$ term renormalizes $\alpha$ by $-4NJ_z\rho^2 d\ell$. Combining the renormalization of $\alpha$ from the $J_{xy}^2$ and $J_z^2$ terms, its overall renormalization is given by
\begin{eqnarray}
d\alpha &=&  4NJ_{xy}^2 \rho^2 d\ell (1-3\alpha)-4N J_z^2 \rho^2 d\ell.
\label{eq.rg2}
\end{eqnarray}

\noindent
Finally, for the coefficient $u_3$ of the $J_{xy}J_z$ term we obtain
\begin{eqnarray}
u_3  &=&
2\left(1-\frac{\alpha}{2}\right) \rho d\ell  \sum_n\sum_{\bk,\bk'}^<
\int d\tau \;\nn\\
 & & \left( c^{\dagger}_{\bk' n\downarrow} (\tau)c_{\bk n\uparrow}(\tau)S^+(\tau)  + {\rm h.c.}\right),
\end{eqnarray}
corresponding to a renormalization of the  $J_{xy}$ coupling, 
\begin{eqnarray}
d J_{xy}= 2\rho d\ell J_{xy}J_z \left(1-\frac{\alpha}{2} \right).
\label{eq.rg3}
\end{eqnarray}

\subsection{Rescaling}

In the neighborhood of the Fermi surface defined by $|\epsilon_{\mathbf{k}}|<\Lambda$, we can approximate the sum over $\mathbf{k}$ states as
\begin{eqnarray}
\sum_{\mathbf{k}} \propto \int d k_{\perp} d\mathbf{k}_{\parallel},
\end{eqnarray}
where $d k_{\perp}$ and $d \mathbf{k}_{\parallel}$ correspond to local changes in components of $\mathbf{k}$ perpendicular and parallel to the Fermi surface, respectively. In particular, $k_{\perp}$ measures the perpendicular distance to the Fermi surface in $\bk$-space. We will also assume a constant density of states throughout the Fermi surface neighbourhood, as well as a linear energy dispersion $\epsilon_{\mathbf{k}}\propto k_{\perp}$. With the above assumptions in mind, the following rescaling will restore the energy cutoff, which has been reduced to $\Lambda e^{-d\ell}$ by integrating out the fast modes
\begin{eqnarray}
k_{\perp} &\rightarrow& e^{-d\ell}k_{\perp}
\nonumber\\
\tau &\rightarrow& e^{d\ell} \tau
\nonumber\\
c^{\dagger}_{n\sigma}(k_{\perp},\mathbf{k}_{\parallel},\tau) &\rightarrow& e^{d\ell/2} 
c^{\dagger}_{n\sigma}(k_{\perp},\mathbf{k}_{\parallel},\tau), 
\label{eq: rescaling}
\end{eqnarray}
The only couplings that will acquire naive scaling as a result are
\begin{eqnarray}
\alpha &\rightarrow& e^{d\ell} \alpha,
\nonumber\\
\beta &\rightarrow& e^{-d\ell}\beta.
\end{eqnarray}

\subsection{RG Equations}

Combining the rescaling and the perturbative corrections (\ref{eq.rg1}), (\ref{eq.rg2}) and (\ref{eq.rg3}), we obtain the following RG equations

\begin{eqnarray}
\label{eq.RGfull}
\frac{d g_z}{d\ell}&=&
g_{xy}^2\left\{
    \begin{array}{ll}
    (1+\alpha) & \mbox{for }S=1,\\ 
    \nn\\
    (1+\alpha)  \left[1 - \alpha \left(1-2S(S+1)\right) \right]  & \mbox{for }S>1,
    \end{array} \right.\nn\\
\frac{d g_{xy}}{d\ell} &=& g_{xy}g_z \left(1-\frac{\alpha}{2} \right),
\nonumber\\
\frac{d \alpha}{d\ell} &=& \alpha + N\left(g_{xy}^2- g_z^2 \right)  - 3N g_{xy}^2 \alpha,
\end{eqnarray}
where we have introduced the dimensionless couplings $g_{xy} := 2J_{xy}\rho$ and $g_z:=2J_z\rho$. For $N=1$ and $S=1$ the RG equations reduce to those given in Eqs. (\ref{eq.RG}).
It is important to note that while for $S=1$ the RG remains self contained, for $S>1$ additional terms that are higher power in the spin operators are generated. These were not present in the original 
anisotropic Kondo model and have been neglected. In a more complete treatment one might include such terms and follow the renormalization of the full crystal-field Hamiltonian in terms of Stevens operators.

\section{2nd Order Perturbation Theory}
\label{ap.2ndorder}

In this Appendix we provide details of the perturbative expansion of the magnetic impurity susceptibilities. To 2nd order in the Kondo couplings the susceptibilities have the general form
\begin{eqnarray}
\chi^{\rm imp}_{\gamma} = \chi_{\gamma}^{\rm free} + \sum_{\gamma'} g_{\gamma'}^2 F_{\gamma\gamma'},
\end{eqnarray}
where $\chi^{\rm free}_\gamma(T,\alpha)$ are the susceptibilities of a free spin subject to single-ion anisotropy $\alpha$. For general spin $S$ we obtain 
\begin{eqnarray}
\chi^{\rm free}_{x,y} &=& \sum_{m=-S}^{S} \frac{e^{- \alpha \Lambda m^2/T}}{4Z^{\rm imp}} \\
& & \times  \Big[ \left(1-e^{-\alpha\Lambda(1-2m)/T}\right) \frac{S(S+1)-m^2+m}{\alpha\Lambda(1-2m)}\nonumber\\
&&+\left(1-e^{-\alpha\Lambda(1+2m)/T}\right)
\frac{S(S+1)-m^2-m}{\alpha\Lambda(1+2m)}
\Big{]},\nn\\
\chi^{\rm free}_{z} &=&
\sum_{m=-S}^{S}  \frac{m^2e^{-\alpha\Lambda m^2/T}}{TZ^{\rm imp}} ,
\end{eqnarray}
where
\begin{equation}
Z^{\rm imp} = \sum_{m=-S}^S e^{-\alpha\Lambda m^2/T}.
\end{equation}

We start by calculating $\chi^{\rm imp}_z$ which requires the evaluation of $F_{zx}=F_{zy}$ and $F_{zz}$. We obtain
\begin{eqnarray}
F_{zx}=F_{zy} &=&\frac{1}{4} \sum_{\bk,\bk',\bq,\bq'}\sum_{n,m=1}^N \int d\tau_1 d\tau_2 d\tau \;\nn\\
& & \Big{\langle} \hat{T}\left( \hat{S}^+(\tau_1) \hat{c}^{\dagger} _{ \mathbf{k}n \downarrow} (\tau_1) \hat{c}_{ \bk' n \uparrow} (\tau_1) +{\rm h.c.} \right)\nn\\
& & \times\left( \hat{S}^+(\tau_2) \hat{c}^{\dagger} _{ \mathbf{q}' m \downarrow} (\tau_2) \hat{c}_{ \bq m \uparrow} (\tau_2) 
+{\rm h.c.} \right)
\nonumber
\\
&&\times
\left( \hat{S}^z(\tau)\hat{S}^z(0) - \left\langle \hat{S}^z(\tau)\hat{S}^z(0) 
\right\rangle  \right)
\Big{\rangle}
\nonumber\\
&=&\frac{1}{2} \int d\tau_1 d\tau_2 d\tau \Big{\langle}\hat{T}\left( \hat{S}^z(\tau)\hat{S}^z(0) - T\chi_z^{\rm free}  \right)\nn\\
& & \times\hat{S}^+(\tau_1) \hat{S}^-(\tau_2) C(\tau_1-\tau_2)\Big{\rangle},
\end{eqnarray}
where the expectation values are taken with respect to $\hat{H}_0$, $\hat{T}$ is the time-ordering operator, and
\begin{eqnarray}
C(\tau_1-\tau_2) & = &  \sum_{\bk,\bk',\bq,\bq'}\sum_{n,m=1}^N\left\langle \hat{T}\hat{c}^{ \dagger}_{\mathbf{k} n\downarrow  } (\tau_1) \hat{c}_{\bq' m \downarrow } (\tau_2) \right\rangle\nn\\
& & \times\left\langle \hat{T}
\hat{c}^{} _{\mathbf{k}' n \uparrow } (\tau_1) \hat{c}^{\dagger}_{ \bq m \uparrow} (\tau_2) 
\right\rangle\nn\\
& = & N\sum_{\mathbf{k},\mathbf{q}} n(-\epsilon_{\bk})n(\epsilon_{\bq})
e^{-|\tau_1-\tau_2|(\epsilon_{\bk} - \epsilon_{\bq})},
\nonumber\\
\end{eqnarray}
where $n(\epsilon_{\bq})=\left(1+e^{\beta\epsilon_{\bq} }\right)^{-1}$. Splitting the integration into two regions, $\tau_1>\tau_2$ and $\tau_1<\tau_2$, and integrating over $\tau$ first, 
we obtain 

\begin{eqnarray}
F_{z,x/y} &=& \frac{N}{8Z^{\rm imp}} \sum_{m=-S}^S  \Big\{ e^{-\alpha\Lambda m^2/T}\left[ S\left(S+1\right) - m^2 + m\right]\nn\\
& & \times\left[ \left( \frac{m^2}{T} - \chi^{\rm free}_z\right)A_{2m-1,1-2m}- m B_{1-2m}\right] 
\nonumber\\
&&+\left[ S\left(S+1\right) - m^2 - m\right]\nn\\
& & \times\left[ \left( \frac{m^2}{T} -\chi_z^{\rm free}\right)
A_{-2m-1,2m+1} + m B_{1+2m}\right]\Big\},\nn\\
\end{eqnarray}
where for $m\neq -n$ 
\begin{eqnarray}
A_{nm} &=&  \frac{Te^{-m\alpha\Lambda /T}}{\alpha\Lambda(n+m)} 
\left[ \left( e^{-n\alpha\Lambda/T} - 1\right)F\left(\frac{n\alpha\Lambda }{T}\right)\right.\nn\\
& & \left. + \left( e^{m\alpha\Lambda/T} - 1\right)F\left(\frac{m\alpha\Lambda }{T}\right)
\right],
\end{eqnarray}
and
\begin{equation}
A_{n,-n} = \left(1-e^{n\alpha\Lambda/T}\right)F'\left(\frac{n\alpha\Lambda}{T}\right) - F\left(\frac{n\alpha\Lambda}{T}\right).
\end{equation}
For the function $B_n$ we obtain
\begin{eqnarray}
B_n &=& \frac{1}{T} \left[
F''\left(\frac{n\alpha\Lambda}{T}\right) \left(e^{-n\alpha\Lambda/T} -1 \right)\right.\nn\\
& & \left. - F'\left( \frac{n\alpha\Lambda}{T}\right) \left( e^{-n\alpha\Lambda/T} +1\right)
\right].
\end{eqnarray}

\noindent
The functions $A_{nm}$ and $B_n$ depend on the integral 
\begin{equation}
F(t) = \int_{-\beta\Lambda}^{\beta\Lambda} dx \int_{-\beta\Lambda}^{\beta\Lambda} dy  \frac{f(x) f(y)}{x+y+t},
\end{equation}
and its derivatives $F'(t)$ and $F''(t)$. Here $f(x) = (1+e^x)^{-1}$ is the Fermi function and in all cases it is understood that the singular contribution from integrating over a narrow region around $x+y+t=0$ is 
excluded, as all singular contributions cancel anyway in the perturbative expansion of $\chi^{\rm imp}_{\gamma}.$ In the case of the integral $F(t)$, this corresponds to simply taking the Cauchy principal value. In the limit 
$\beta\Lambda\rightarrow\infty$, we can approximate the integral $F(t)$ as follows,

\begin{eqnarray}
F(t) &\approx& -2\beta\Lambda\log 2 - t\log\frac{\beta\Lambda}{2}\nn\\
& & +\int_{-\infty}^{\infty} dx \int_{-\infty}^{\infty} dy\;
(x+y+t) \nn\\
& & \times  \log \left | \frac{x+y+t}{e} \right |
\; f'(x) f'(y).
\label{eq: approximated Kondo integrals1}
\end{eqnarray}

\noindent
In this approximation, the derivatives are given by 
\begin{eqnarray}
F'(t) &\approx& -\log \frac{\beta\Lambda}{2} 
+ \int_{-\infty}^{\infty} dx \int_{-\infty}^{\infty} dy\;
 \log \left | x+y+t \right | \nn\\ 
 & & \times f'(x) f'(y), \nn\\
F''(t) &\approx&
\int_{-\infty}^{\infty} dx \int_{-\infty}^{\infty} dy\;
(x+y+t) \nn\\ 
& & \times \log \left | \frac{x+y+t}{e} \right |
\; f''(x) f''(y). \label{eq: approximated Kondo integrals2}
\end{eqnarray}

\begin{figure}
\centering
\includegraphics[width=1.0\linewidth]{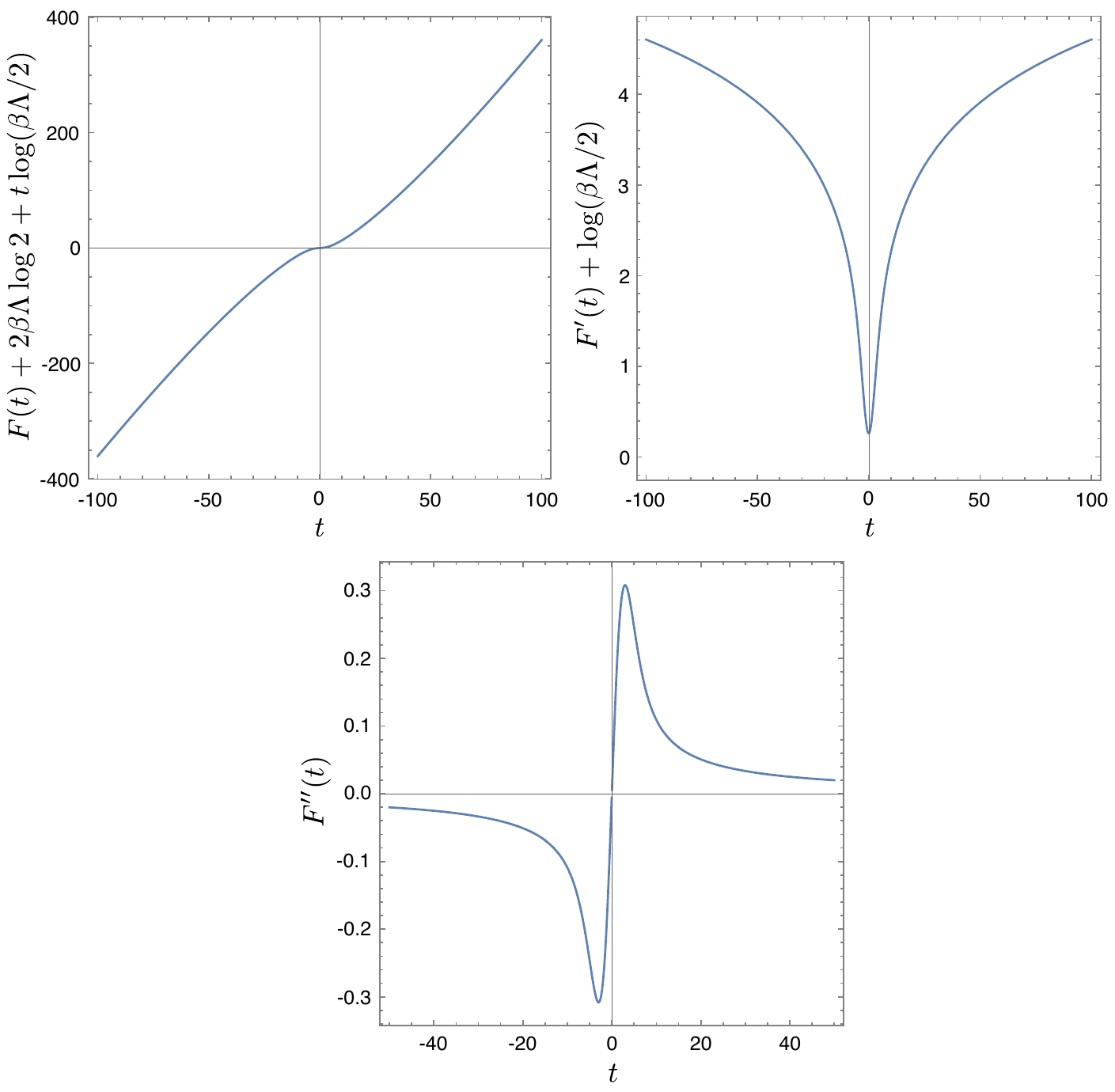}
\caption{The figure shows the behaviour of the three integrals contained in the definitions of $F(t)$ and its derivatives in the limit $\beta\Lambda\rightarrow\infty$, that are given in 
Eqs.~(\ref{eq: approximated Kondo integrals1}) and (\ref{eq: approximated Kondo integrals2}). The asymptotic behaviour, as $|t|\rightarrow\infty$, of the three integrals contained in $F(t)$, $F'(t)$ and $F''(t)$ is given by $t\log\left( |t|/e\right)$,  $\log |t|$ and $1/t$ respectively.   }
\label{figure7}
\end{figure}

Fig.~\ref{figure7} shows the behaviour of the approximated integrals for $F(t)$, $F'(t)$ and $F''(t)$. Note that whenever we present results for high temperatures $T\gtrsim \Lambda$, the non-approximated 
versions of $F(t)$ and its derivatives are used.

\noindent
For the expansion coefficient $F_{zz}$ we simply obtain 
\begin{eqnarray}
F_{zz} & = & \beta \int d\tau_1 d\tau_2 \left( \left\langle
( \hat{S}^{z})^4 \right\rangle
 - \left\langle (\hat{S}^{z})^2 
\right\rangle^2\right)  C(\tau_1-\tau_2)
\nonumber\\
&=&
-\frac{F(0)}{2 T Z^{\rm imp}} 
\left[ \sum_m m^4 e^{-m^2\alpha\Lambda/T}
-\left(T\chi_z^{\rm free}\right)^2
\right].
\end{eqnarray}

In order to compute the  perturbative corrections to $\chi^{\rm imp}_x = \chi^{\rm imp}_y$ we need to compute the coefficients 
$F_{xz}$ and $F_{xx}=F_{xy}$. The resulting expressions are quite lengthy and to the best of our knowledge have not been calculated before. The 
coefficients can be written as
\begin{eqnarray}
F_{xz} &=&  \int d\tau_1 d\tau_2 d\tau \Big\langle\hat{T}\left( \hat{S}^x(\tau)\hat{S}^x(0) - T\chi^{\rm free}_x \right)\nn\\
& & \times \hat{S}^z(\tau_1) \hat{S}^z(\tau_2) C(\tau_1-\tau_2)
\Big\rangle,\\
F_{xx} &=& \frac{1}{4} \int d\tau_1 d\tau_2 d\tau \Big\langle \hat{T}\left( \hat{S}^x(\tau)\hat{S}^x(0) - T\chi^{\rm free}_x  \right)\nn\\
& & \times\left(
\hat{S}^+(\tau_1) \hat{S}^-(\tau_2) + {\rm h.c.}
\right) C(\tau_1-\tau_2)
\Big\rangle.
\end{eqnarray}

We take $\tau_2>\tau_1$ (by symmetry equivalent to the region $\tau_2<\tau_1$) and split the integration over $\tau$ into three regions depending on its position relative to $\tau_1,\tau_2$. We finally 
integrate over $\tau_1,\tau_2$ to obtain
\begin{eqnarray}
F_{xz} &=& \frac{N}{8\alpha\Lambda Z^{\rm imp}}\sum_m e^{-\alpha\Lambda m^2/T} \Big\{
 \nn\\
& & \kappa(-m) \Big[ \frac{m^2}{1-2m}\left(A_{0,0} - A_{1-2m,0} \right)\nn\\
& & +\frac{(m-1)^2}{1-2m} \left(A_{0,1-2m} - A_{0,0} e^{-(1-2m)\alpha\Lambda/T}\right)\nn\\
& & + \frac{m(m-1)}{1-2m} \left( A_{1-2m,0} -A_{0,1-2m} \right) \Big]\nn\\
&&+\kappa(m) \Big[\frac{m^2}{1+2m}\left(A_{0,0} - A_{1+2m,0} \right)\nn\\
& & +\frac{m(m+1)}{1-2m} \left( A_{1+2m,0} -A_{0,1+2m} \right)\nn\\
&&+\frac{(m+1)^2}{1+2m} \left(A_{0,1+2m} - A_{0,0} e^{-(1+2m)\alpha\Lambda/T}\right) \Big]\Big\}\nn\\
& & -\frac{NT}{2} \chi_x^{\rm free} \chi_z^{\rm free} A_{0,0},
\end{eqnarray}
where we have defined $\kappa(m) = S(S+1)-m(m+1)$, for brevity. A similar calculation gives

\begin{eqnarray}
F_{xx} &=&
\frac{N}{16\alpha\Lambda Z^{\rm imp}}
\sum_m e^{-\alpha\Lambda m^2/T} 
\Big\{\nn\\
& & \kappa^2(m) \frac{A_{-1-2m,1+2m}-A_{0,1+2m}}{1+2m}
\nonumber\\
&&+
\kappa(m)\kappa(-m)
\frac{A_{-1-2m,1+2m}-A_{-4m,1+2m}}{1-2m}
\nonumber\\
&&+
\kappa(m)\kappa(-m)
\frac{A_{2m-1,1-2m} - A_{4m,1-2m}}{1+2m}
\nonumber\\
&&+
\kappa^2(-m)
\frac{A_{2m-1,1-2m}-A_{0,1-2m}}{1-2m}
\nonumber\\
&&+
\kappa(m)\kappa(m+1)
\frac{A_{0,1+2m} - A_{-3-2m,4+4m}}{3+2m}
\nonumber\\
&&+
\kappa(m)\kappa(-m)
\frac{A_{1-2m,0}-A_{-4m,1+2m}}{1+2m}
\nonumber\\
&&+
\kappa(-m)\kappa(-m+1)
\frac{A_{0,1-2m} - A_{2m-3,4-4m} }{3-2m}
\nonumber\\
&&+
\kappa(m)\kappa(-m)
\frac{A_{4m,1-2m}-A_{1+2m,0}}{2m-1}
\nonumber\\
&&+
\kappa(m)\kappa(m+1)\nn\\
& & \times \frac{A_{-2m-3,4+4m} - A_{-2m-3,2m+3}e^{-(1+2m)\alpha\Lambda/T}}{1+2m}
\nonumber\\
&&+
\kappa^2(-m)
\frac{A_{1-2m,0} - A_{1-2m,2m-1}e^{(1-2m)\alpha\Lambda/T}}{1-2m}
\nonumber\\
&&+
\kappa(-m)\kappa(-m+1)\nn\\
& & \times \frac{A_{2m-3,4-4m} -A_{2m-3,3-2m}e^{-(1-2m)\alpha\Lambda/T} }{1-2m}
\nonumber\\
&&+
\kappa^2(m)
\frac{A_{1+2m,0} - A_{1+2m,-1-2m}e^{-(1+2m)\alpha\Lambda/T} }{1+2m}
\nonumber\\
&&
-4\alpha\Lambda\chi_x^{\rm free}\kappa(m)
A_{-2m-1,1+2m}\nn\\
& & -4\alpha\Lambda\chi_x^{\rm free}
\kappa(-m)
A_{2m-1,1-2m}\Big\},
\end{eqnarray}

In this section, we have derived analytic expressions for the impurity susceptibilities $\chi^{\rm imp}_z$ and $\chi^{\rm imp}_x=\chi^{\rm imp}_y$ to second order in the Kondo couplings $g_z$, $g_{xy}$ and for general 
values of $N$, $S$ and the single-ion anisotropy $\alpha$. In the isotropic limit $\alpha = 0$, $g_{xy}=g_z=g$, and letting $\beta\Lambda\rightarrow\infty$, we recover the following analytic result
\begin{eqnarray}
\chi^{\rm imp}_{\gamma} &=& \chi_{\gamma}^{\rm free} \left[ 1 +Ng^2 F'(0) \right]\nn\\
& = &  \frac{S(S+1)}{3T}\left[ 1 -Ng^2 \log \left(C\beta\Lambda \right)\right] ,
\label{eq: analytic suscep}
\end{eqnarray}
where $C=0.385$. This analytic result was used to benchmark our calculations.

\begin{figure}
\centering
\includegraphics[width=\columnwidth]{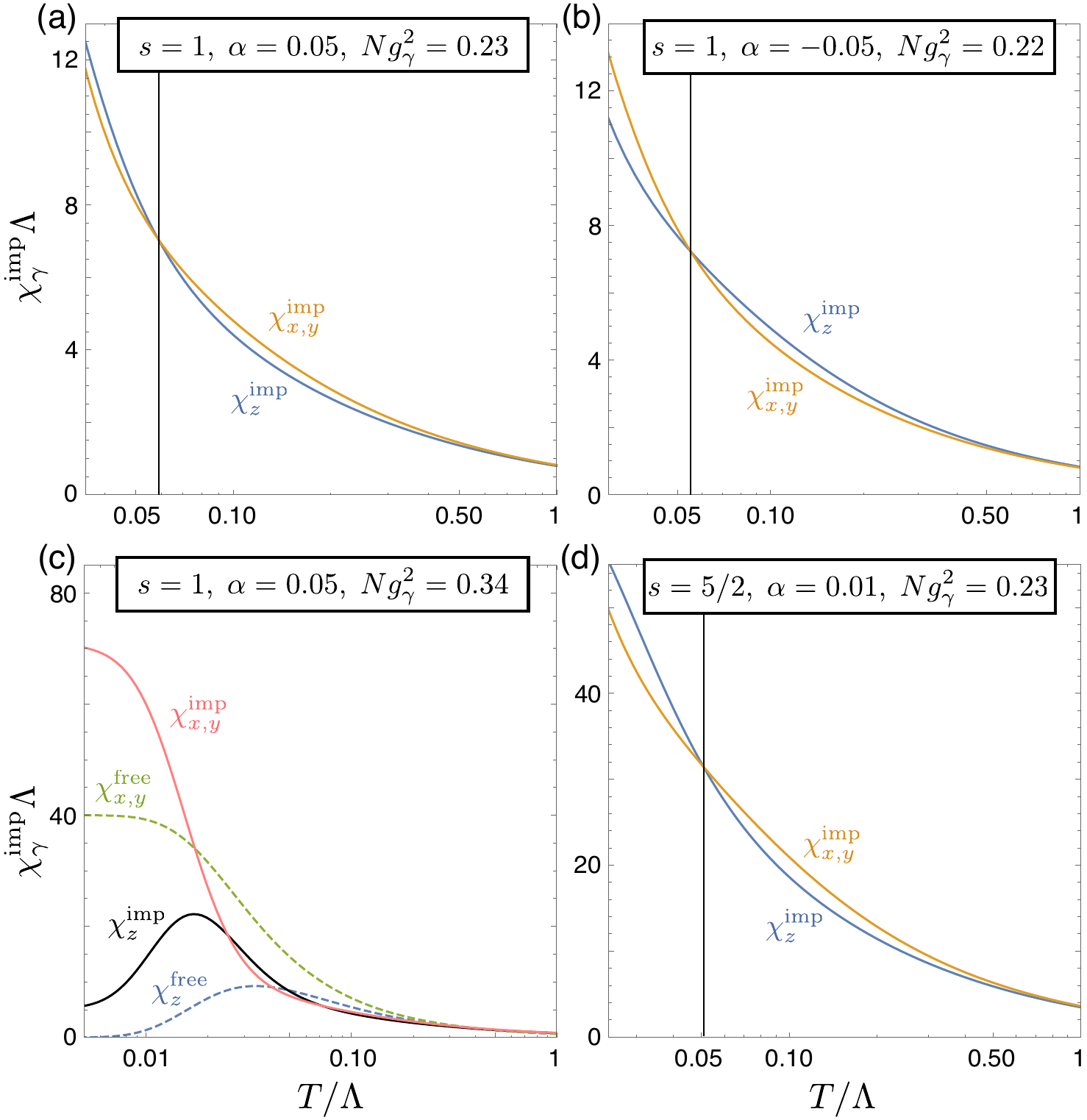}
\caption{The crossing of the susceptibilities $\chi^{\rm imp}_z$ and $\chi^{\rm imp}_{x,y}$ in second order perturbation theory for different choices of the parameters $S,N,g_{\gamma},\alpha$:
(a) $S=1$  with easy-plane anisotropy $\alpha=0.05$ and isotropic Kondo interaction $Ng_{\gamma}^2=0.23$. Note that the crossing takes place at a temperature, where the Kondo corrections to $\chi^{\rm free}_z$ are about $10\%$ and to $\chi^{\rm free}_{x,y}$ are about $40\%$.
(b) $S=1$ with easy-axis anisotropy $\alpha=-0.05$ and isotropic Kondo interaction $Ng_{\gamma}^2=0.22$. 
(c) $S=1$ with easy-plane anisotropy $\alpha=0.05$ and isotropic Kondo interaction  $Ng_{\gamma}^2=0.34$. The difference between $\chi^{\rm imp}_z$ and $\chi^{\rm imp}_{x,y}$ is much smaller than the difference between $\chi^{\rm free}_z$ and $\chi^{\rm free}_{x,y}$  above the crossing temperature of $T\approx 0.07\Lambda$, indicating a reduction of the effective positive $\alpha$ by Kondo screening, in line with RG results. The effective $\alpha$ turns negative when $\chi_z^{\rm imp}>\chi_{x,y}^{\rm imp}$, which also agrees with RG analysis. Note that $\chi_z^{\rm imp}$ remains greater than $\chi_{x,y}^{\rm imp}$ for temperatures as low as half of the splitting between the $S^z$ eigenstates, where there is a second crossing, and note also that $\chi_z^{\rm imp} \rightarrow \frac{Ng_{xy}^2}{\alpha\Lambda} \neq 0$ as $T\rightarrow 0$. This second crossing and the low-temperature limit are not observed in RG, which takes into account feedback from higher order terms.
(d) $S=5/2$ in the presence of easy-plane anisotropy $\alpha=0.01$ and isotropic Kondo interaction of $Ng_{\gamma}^2=0.23$.}
\label{figure8}
\end{figure}

Fig.~\ref{figure8} shows a representative sample of our results.
In line with our RG results, no crossing  of the susceptibilities $\chi_{x,y}^{\rm imp}$ and $\chi_z^{\rm imp}$ takes place for large $\alpha$, regardless of the strength of the isotropic Kondo coupling $g=g_{xy}=g_z$. Similarly, in line with our RG analysis, for small anisotropy $\alpha$, crossing only takes place above a critical Kondo coupling. We also note here that the crossing of susceptibilities can be observed for any $S\geq 1$ and any $N$, over a particular range of parameters $\alpha,g_{\gamma}$.

\section{Details on NRG calculation}
\label{ap.NRG}

In this Appendix, we summarise our particular implementation of the NRG routine to calculating the susceptibility of the Kondo model presented in Figs.~\ref{figure3} and \ref{figure4}.  In the NRG method, the continuum conduction band is logarithmically discretised in energy and the electronic part of the  Hamiltonian is mapped to a semi-infinite chain with exponentially decaying hopping, with the impurity placed at the zeroth site. We start off  with the impurity coupled to a single electronic site via the Kondo exchange and the Hamiltonian is then diagonalised iteratively. At every iteration: (i) a new site of the chain is added to the Hamiltonian, (ii) the new Hamiltonian is then diagonalised, (iii) high energy states are thrown away. In this way, we zoom into lower and lower energy scales that are relevant at low temperatures. Details of the NRG method can be found in the comprehensive reviews of Ref.~\cite{Krishna-Murthy+80} and Ref.~\cite{Costi+08}.

Ref.~\cite{Zitko+08} used NRG to calculate the total susceptibility of Kondo models with easy-axis and easy-plane anisotropies but only in the $z$-direction, where the corresponding magnetisation commutes with the Hamiltonian. This susceptibility is a thermodynamic observable that can be computed using the fluctuation-dissipation theorem
\begin{eqnarray}
\label{eq.38}
\displaystyle{\chi_z=\frac{1}{T(n)} \sum_{i(n)}  e^{-E_{i(n)}  / T(n)}\langle i(n) | (\hat{J}^z)^2 | i(n) \rangle },
\end{eqnarray}
where $\hat{J}^z=\hat{S}^z+\hat{s}^z$ is the total angular momentum operator in the z-direction and $i(n)$ indexes the energy eigenstates at the $n$-th NRG iteration. The energy scale at the $n$-th iteration sets the temperature at which the observables will be worked out most accurately
\begin{eqnarray}
T(n)=\Lambda^{-(n-1)/2}.
\end{eqnarray}

\noindent
We have extended the calculation of Ref.~\cite{Zitko+08} to other directions by computing the susceptibility in the $x$-direction 
\begin{eqnarray}
\label{eq.40}
\chi_x & = &  \sum_{i(n), j(n)} \frac{e^{-E_{i(n)}/T(n)}- e^{-E_{j(n)}/T(n)} }{E_{j(n)} - E_{i(n)}} \nonumber\\
& & \times\left| \langle i(n) |    \hat{J}^x  
|j(n) \rangle \right|^2,
\end{eqnarray}
where the appropriate limit needs to be taken in case of degenerate eigenstates $E_{i(n)}=E_{j(n)}$.
We emphasise that, unlike $\chi_z$, $\chi_x$ is a dynamical observable that is sensitive to a broad range of energy scales, not just energies of order $T(n)$, and an accurate calculation requires more eigenstates to be kept at each NRG iteration. We refer to the reader to the excellent review of Ref.~\cite{Costi+08} for a detailed discussion of the difficulties associated with calculating dynamical observables. To maximise the number of eigenstates that can be kept at each iteration, we made use of conserved observables: charge $q$ and total angular momentum $J_z$. The Hamiltonian is block-diagonal with respect to subspaces labelled $J_z$ and $q$, which allowed us to speed up the routine.  We have also used particle-hole symmetry and spin reflection symmetry to analytically relate degenerate energy eigenstates with charges $2n_s-q \leftrightarrow q$ and total angular momenta  $J^z\leftrightarrow -J^z$ through the following unitary transformations that commute with the Hamiltonian
\begin{eqnarray}
|   2n_s - q \rangle &=& \exp \left[ -\frac{\pi}{2}\sum_{i=1}^{n_s} \left( c_{i\downarrow} c_{i\uparrow}  -c^{\dagger}_{i\uparrow} c^{\dagger}_{i\downarrow} 
\right)\right] | q \rangle,
\\
|  -J_z  \rangle &=& \exp \left[-i\pi \hat{S}^y - \frac{\pi}{2}\sum_{i=1}^{n_s} \left( c^{\dagger}_{i\uparrow} c_{i\downarrow} - c^{\dagger}_{i\downarrow} c_{i\uparrow} 
\right)\right]    | J_z \rangle,\nn
\end{eqnarray}
where $n_s$ is the number of electronic sites in the semi-infinite chain, which are indexed by $i$.
We have thus been able to decrease the number of eigenstates that need to be independently parametrised and stored by a factor of four, allowing us to  keep more eigenstates in the NRG routine.

We benchmarked our NRG calculations by comparing our results for $\chi_z$ against those of Ref.~\cite{Zitko+08}. In order to do this accurately, we have used the same parameters
\begin{eqnarray}
\Lambda &=&2,\nn
\\
g&=&2 J\rho=0.1,
\\
T_{\rm K} &=& W\sqrt{g} e^{-1/g}=1.4\times 10^{-5} W \frac{}{},\nn
\end{eqnarray}
where $\Lambda$ is the NRG discretisation parameter and $T_{\rm K}$ the Kondo temperature (Note that our definition of $J$ is half the value of the corresponding definition used in Ref.~\cite{Zitko+08}). Similarly to this work, we have ensured that the truncation energy is around $10T(n)$ and the energy gap is at least $0.01T(n)$ at the point of truncation. There is excellent agreement between our NRG results for $\chi_z$, given in Fig.~3  and those presented in Ref.~\cite{Zitko+08}.


\begin{thebibliography}{64}
\bibitem [{per()}]{pert}%
  \BibitemOpen
  \href@noop {} {}\bibinfo {howpublished} {Moreover, they pointed out that the
  calculation of $\chi^{\rm imp}_{x,y}$ would require a much larger amount of
  algebraic work, owing to the non-commutativity between the single-ion
  anisotropy and the $J_{x,y}$ terms.}\BibitemShut {Stop}%
\end{thebibliography}
\end{document}